\documentclass[epj,nopacs]{svjour}

\usepackage[stretch=10,shrink=10]{microtype}

\usepackage{amsmath}
\usepackage{amssymb}

\usepackage{graphicx}
\usepackage[bookmarks=true,colorlinks,linkcolor=blue,urlcolor=blue,citecolor=blue]{hyperref} 
\usepackage[dvipsnames]{xcolor}
\usepackage{standalone}
\usepackage{float}

\usepackage[normalem]{ulem} 
\usepackage{cite}
\usepackage{dblfloatfix}

\makeatletter
\newcommand{\printfontsize}{\f@size pt}
\makeatother

\newcommand{\pdagger}{\phantom{\dagger}}
\newcommand{\diff}{\frac{d}{d\Lambda}}
\renewcommand{\vec}{\boldsymbol} 

\newcommand{\chicw}{\chi_{\mathrm{C}}}  
\newcommand{\Lambdacw}{\Lambda_{\mathrm{C}}}
\newcommand{\Lambdatransition}{\Lambda_{\mathrm{T}}}

\definecolor{darkblue}{HTML}{004D6B}
\definecolor{darkred}{HTML}{8c1515}
\definecolor{darkgreen}{HTML}{006400}
\usepackage{hyperref}
\hypersetup{
	colorlinks=true,
	urlcolor=darkred,
	citecolor=darkblue,
	linkcolor=darkred,
	breaklinks
}
\title{%
    Benchmark Calculations of Multiloop Pseudofermion fRG
}
\author{Marc K.~Ritter\inst{1}, 
		Dominik Kiese\inst{2}, 
		Tobias M\"uller\inst{3},
		Fabian B.~Kugler\inst{4},
		Ronny Thomale\inst{3},
		Simon Trebst\inst{2},\\
		Jan von Delft\inst{1}
		}
\institute{
	 Arnold Sommerfeld Center for Theoretical Physics, Center for NanoScience, and Munich Center for Quantum\\ Science and Technology, Ludwig-Maximilians-Universit\"at M\"unchen, 80333 Munich, Germany
	\and Institute for Theoretical Physics, University of Cologne, 50937 Cologne, Germany
	\and Institute for Theoretical Physics, University of W\"urzburg, Am Hubland, D-97074 W\"urzburg, Germany
	\and Department of Physics and Astronomy, Rutgers University, Piscataway, New Jersey 08854, USA
}
\date{\today}

\abstract{%
The pseudofermion functional renormalization group (pffRG) is a computational method for determining zero-temperature phase diagrams of frustrated quantum magnets. 
In a recent methodological advance, the commonly employed Katanin truncation of the flow equations was extended to include multiloop corrections, thereby capturing additional contributions from the three-particle vertex \cite{thoenniss2020,kiese2021}.
This development has also stimulated significant progress in the numerical implementation of pffRG, allowing one to track the evolution of pseudofermion vertices under the renormalization group flow with unprecedented accuracy. However, cutting-edge solvers differ in their integration algorithms, heuristics to discretize Matsubara frequency grids, and more. To lend confidence in the numerical robustness of state-of-the-art multiloop pffRG codes, we present and compare results produced with two independently developed and algorithmically distinct solvers for Heisenberg models on three-dimensional lattice geometries. Using the cubic lattice Heisenberg (anti)ferromagnet with nearest and next-nearest neighbor interactions as a generic benchmark model, we find the two codes to quantitatively agree, often up to several orders of magnitude in digital precision, both on the level of spin-spin correlation functions and renormalized fermionic vertices for varying loop orders. 
These benchmark calculations further substantiate the usage of multiloop pffRG solvers to tackle unconventional forms of quantum magnetism. 
}

\begin{document}
\maketitle


\section{Introduction}

A fascinating phenomenon in the study of frustrated quantum magnets is the interplay of unconventional forms of magnetic order and the possible emergence of quantum spin liquid states near zero temperature \cite{lacroix2011introduction}.
The successful description of such low-energy states of quantum spin systems has, however, remained challenging, especially in the presence of competing interactions, geometric frustration, and in higher spatial dimensions.

Since its inception more than a decade ago \cite{ReutherOrig}, the pseudofermion functional renormalization group (pffRG) has become a powerful and flexible approach to map out the zero-temperature phase diagrams of various quantum spin models, both in two \cite{ReutherOrig,Reuther-2011a,Reuther-2011b,Reuther-2011c,ReutherKitaev,Reuther-2012,Reuther-2014a,Suttner-2014,Reuther-2014b,Iqbal-2015,Iqbal-2016b,Iqbal-2016a,Buessen-2016,Keles-2018a,Keles-2018b,KieseSpinValley,astrakhantsev2021pinwheel} and three spatial dimensions \cite{Iqbal3D,Buessen-2016,Iqbal-2017,BuessenDiamond,Iqbal-2018a,MuellerPyrochlore,MuellerBCC,KieseFCC,Ghosh-2019,Chillal-2020}. Although the problem obtained after representing the spin operators by complex fermions is treated approximately, one of the striking features of pffRG is its ability to track competing instabilities in different interaction channels, allowing one to discriminate putative spin-liquid phases from long-range ordered magnetic ground states. This ability can be traced back \cite{LargeS,LargeN} to the inclusion of leading-order $1/S$ and $1/N$ diagrams (the former promoting classical magnetic order, the latter quantum fluctuations), which are treated on equal footing in pffRG by means of the routinely employed Katanin truncation \cite{Katanin2004}.

Recently, the multiloop truncation scheme of the infinite hierarchy of fRG flow equations \cite{Kugler_1,Kugler_2,Kugler_3},
previously used in the context of the Hubbard \cite{Tagliavini2019,Hille2020}
and Anderson impurity model \cite{Chalupa2021},
was applied to the zero-temperature pffRG by some of us \cite{thoenniss2020,kiese2021}. The convergence in the number of loops over a wide range of energy scales attested to the inner consistency of the pffRG method, despite being used in the strong-coupling limit. These developments were accompanied and facilitated by substantial improvements of the numerical implementation that remedy many shortcomings of previous studies. Yet, some of these advances, such as the employed integration routines and adaptive Matsubara frequency grids \cite{thoenniss2020,kiese2021}, rely on certain numerical heuristics, affecting, e.g., the minimal grid spacing and largest Matsubara frequencies considered. Therefore, quantitative agreement between different implementations is, although highly desired, not guaranteed {\sl a priori}.

In the present work, we provide evidence for the numerical robustness of pffRG by benchmarking two independent state-of-the-art solvers, one provided by a research group at LMU Munich (dubbed code \#1 in the following), and one by a Cologne--W\"{u}rzburg collaboration (denoted by code \#2) with an open-source release \cite{PFFRGjl}. As a test case, we consider ferro- and antiferromagnetic Heisenberg models on the simple cubic lattice and compare our results both on the level of renormalized couplings (i.e.\ fermionic vertex functions) as well as for the (post-processed) spin-spin correlation functions.

The remainder of the paper is structured as follows. We begin by providing a brief overview of the multiloop pffRG in Sec.~\ref{sec:mfRG}. This is followed by an in-depth comparison of the numerical results produced by the two codes at hand in Sec.~\ref{sec:results}. Finally, in Sec.~\ref{sec:technical_aspects}, technical aspects of the implementation, such as the choice of frequency grids, integration routines and differential equation solvers are discussed, with special emphasis devoted to their influence on the numerical stability and accuracy of the two codes.


\section{Multiloop pseudofermion fRG}
\label{sec:mfRG}

Within the pffRG approach, one can study generic spin-$1/2$ Hamiltonians with bilinear spin couplings, i.e.,
\begin{align}
	\mathcal{H} = \tfrac{1}{2} \sum_{ij} J^{\mu \nu}_{ij} S^{\mu}_i S^{\nu}_j \,.
	\label{ham}
\end{align}
Here, the spin operators $S^{\mu}_i$ live on the sites $i$ of an arbitrary lattice, and the exchange matrices $J^{\mu \nu}_{ij}$ are assumed to be real. The spin operators are represented in terms of complex pseudofermions $f^{(\dagger)}_{i \alpha}$ with $\alpha \in \{\uparrow, \downarrow \}$ as
\begin{align}
    S^{\mu}_i = \tfrac{1}{2} \sum_{\alpha, \beta} f^{\dagger}_{i \alpha} \sigma^{\mu}_{\alpha \beta} f^{\pdagger}_{i \beta} \,,
    \label{partons}
\end{align}
where $\sigma^{\mu}_{\alpha \beta}$ for $\mu \in \{x, y, z\}$ are the Pauli matrices. This yields a purely quartic Hamiltonian which can be treated by established functional RG techniques. 

Note that the pseudofermion representation of the spin algebra comes with an artificial enlargement of the local Hilbert space dimension, which must be dealt with by an additional particle number constraint $\sum_{\alpha} f^{\dagger}_{i\alpha} f^{\pdagger}_{i \alpha} = 1$ on every lattice site.
In practice, this constraint is not enforced, but holds on average due to particle-hole symmetry \cite{ReutherOrig,kiese2021,thoenniss2020}. Nevertheless, the influence of fluctuations can be quantitatively gauged by explicitly computing the variance of the number operator, which can be expressed through the equal-time spin-spin correlation function $\langle S_{i}^{\mu} S_{i}^{\mu} \rangle$ \cite{thoenniss2020}. Although fluctuations are not fully suppressed, even if a local level repulsion term $A S^{\mu}_i S^{\mu}_i$ (with $A < 0$) is employed, recent studies \cite{thoenniss2020,LargeS,BuessenDiamond,KieseSpinValley} pointed out that observables extracted from pffRG flows are qualitatively unaffected by the unphysical Hilbert space sectors.

An alternate decomposition of the spin operators into Majorana instead of Abrikosov fermions allows one to circumvent the problem of unphysical states in the fermionic representation at the cost of redundant copies of physical Hilbert-space sectors \cite{pmfRG1}. For moderately high temperatures, the latter approach was recently shown to enable an accurate calculation of thermodynamic observables \cite{pmfRG2}, such as the free energy and specific heat. However, the approach was also found to suffer from unphysical divergencies when approaching the $T \to 0$ limit, which we consider here (for the Abrikosov fermion decomposition). 

Since kinetic contributions are absent in the pseudofermion representation of Eq.~\eqref{ham}, the free propagator assumes the simple form
\begin{align}
	G_0(1'|1) &= (i \omega_{1})^{-1} \delta_{i_{1'} i_{1}} \delta_{\alpha_{1'} \alpha_{1}} \delta(\omega_{1'} - \omega_{1}) \,,
\end{align}
diagonal in all indices. In order to successively integrate out high-energy modes and thus provide an effective low-energy description of a given model, a cutoff parameter, here denoted as $\Lambda$, is introduced in the bare propagator. The fRG equations then govern the flow of the $n$-particle vertices from the UV limit $\Lambda \to \infty$, where the regularized bare propagator vanishes, to the infrared limit $\Lambda \to 0$, where one recovers the physical theory. As such, there is a certain degree of freedom in the cutoff implementation. A popular choice for the regulator in pffRG is a Heavyside step function, which sharply suppresses frequency contributions $|\omega| < \Lambda$. This choice is very useful for analytical treatments of pffRG in the large-$S$ and large-$N$ limit, where the flow equations can be solved exactly and reproduce mean-field gap equations \cite{LargeS,LargeN}. However, if numerical calculations are employed away from these limits, a non-analytic regulator spoils the smoothness of the right-hand side of the flow equations, and therefore limits the applicability of higher-order integration routines. For this reason, we consider a smooth regulator
\begin{align}
    R^{\Lambda}(\omega) = 1 - e^{-\omega^2 / \Lambda^2} \,,
    \label{regulator}
\end{align}
throughout this manuscript, and implement the cutoff as $G^{\Lambda}_0(\omega) = R^{\Lambda}(\omega) G_0(\omega)$, with $G_0(\omega) \equiv  (i \omega)^{-1}$.

In order to make the infinite hierarchy of fRG flow equations amenable to further calculations, a truncation is necessary. Usually, this is done by neglecting all $n$-particle vertices of $n=3$ and higher \cite{Katanin2004}.
However, to capture the physics of interest in pffRG,
one must already go beyond that by using the Katanin truncation, which feeds the $\Lambda$ derivative of the self-energy $\Sigma^{\Lambda}$ back into the flow of the two-particle vertex $\Gamma^{\Lambda}$ \cite{ReutherOrig}. Within this truncation, the flow equations schematically read
\begin{align}
	\diff \Sigma^\Lambda &= - \big{[} \Gamma^{\Lambda} \circ S^{\Lambda} \big{]}_{\Sigma} \,,
	\label{self}
	\\
	\diff \Gamma^\Lambda
	& =
	\sum_c \dot{\gamma}_{c}^\Lambda
	=
	-\sum_c [\Gamma^\Lambda \circ \partial_\Lambda (G^\Lambda \times G^\Lambda) \circ \Gamma^\Lambda]_{c} \,.
	\label{eq:bubble_function}
\end{align}
Here, we introduced the loop function $[\Gamma \circ G]_{\Sigma}$ and the single-scale propagator $S^{\Lambda} \equiv -\diff G^{\Lambda}|_{\Sigma^{\Lambda} = \text{const.}}$.
We categorized the contributions to the flow of $\Gamma$ into three distinct channels $c$:
the particle-particle ($s$) channel, the direct particle-hole ($t$) channel, and the crossed particle-hole ($u$) channel. Each ``bubble'' term, with the general form $[\Gamma \circ (G \times G') \circ \Gamma']_c$, describes the flow of a two-particle reducible vertex $\gamma_c$.
As all self-energies, vertices, and related correlators are $\Lambda$-dependent, we refrain from writing this dependence explicitly in the following.

The multiloop fRG (mfRG) flow \cite{Kugler_1,Kugler_2,Kugler_3}, recently employed within pffRG \cite{thoenniss2020,kiese2021}, is an attempt to go beyond the Katanin truncation and capture even more contributions from $n$-particle vertices with $n\geq3$. It can be derived from the parquet approximation \cite{Parquet}, which self-consistently connects one- and two-particle correlation functions via the Schwinger--Dyson (SDE) and Bethe--Salpeter equations (BSE), and as such the inherent dependence of the $\Lambda \to 0$ fRG result on the specific choice of regulator is eliminated \cite{Kugler_2}. This approximation includes all those contributions to the flow of the two-particle vertex which can be efficiently calculated, i.e., with the same cost as the one-loop flow in Eqs.~\eqref{self} and \eqref{eq:bubble_function}. Summarized briefly: To obtain the mfRG flow of $\gamma_c$, one iteratively computes multiloop corrections to the one-loop ($\ell=1$) result, using bubble functions with undifferentiated propagators but differentiated vertices. In a similar fashion, one can recover equivalence to the SDE, by feeding back the so-determined vertex corrections into the self-energy flow. 

One of the most important ingredients to achieve sufficient numerical accuracy throughout the multiloop flow, is an appropriate treatment of the frequency dependence of the two-particle vertex. In Ref.~\cite{WentzellAsymptotics}, a pa\-ra\-me\-tri\-za\-tion in terms of one bosonic and two fermionic frequencies (the fourth frequency argument is fixed by energy conservation) for each two-particle reducible vertex was put forward. This pa\-ra\-me\-tri\-za\-tion captures the non-trivial high frequency asymptotics of the vertices while being numerically efficient. Code \#1 uses precisely the proposal of Ref.~\cite{WentzellAsymptotics}, and the diagrams contributing to each channel are grouped into four asymptotic classes $K_n$ as
\begin{align}
    \gamma_{c}(\omega_c, \nu_c, \nu'_c) &= K_{1, c}(\omega_c) \notag \\ 
                                     &+ K_{2, c}(\omega_c, \nu_c) + K_{2', c}(\omega_c, \nu'_c) \notag \\ 
                                     &+ K_{3, c}(\omega_c, \nu_c, \nu'_c) \,,
\end{align}
where we displayed only frequency arguments for brevity. Here, $\omega_c, \nu_c$ and $\nu'_{c}$, denote the natural frequency arguments for diagrams reducible in channel $c$ (see Ref.~\cite{thoenniss2020} for the conventions used). The $K_{n}$ asymptotically decay to zero in each frequency, allowing one to reduce the necessary number of arguments when summing up the asymptotic classes to obtain $\gamma_c$. Code \#2 chooses a slightly different approach, by defining asymptotic classes $Q_n$ \cite{li_asymptotics} as
\begin{align}
    Q_{1, c}(\omega_c)                &= K_{1, c}(\omega_c) \notag\\
    Q_{2, c}(\omega_c, \nu_c)         &= K_{1, c}(\omega_c) + K_{2, c}(\omega_c, \nu_c)\notag\\
    Q_{2', c}(\omega_c, \nu'_c)  &= K_{1, c}(\omega_c) + K_{2', c}(\omega_c, \nu'_c)\notag\\
    Q_{3, c}(\omega_c, \nu_c, \nu'_c) &= K_{1, c}(\omega_c) \notag \\ 
                                      &+ K_{2, c}(\omega_c, \nu_c) + K_{2', c}(\omega_c, \nu'_c) \notag \\ 
                                      &+ K_{3c}(\omega_c, \nu_c, \nu'_c) \,,
                                      \label{Qdef}
\end{align}
with the respective choice of natural frequency arguments outlined in Ref.~\cite{kiese2021}. Since the $K_n$ classes decay to zero for large frequencies, the $Q_n$ (at least for $n > 1$) are projected to a lower class. For instance, $Q_{3, c}(\omega_c, \nu_c, \nu'_c) = Q_{2, c}(\omega_c, \nu_c)$ if $|\nu'_c| \to \infty$.
Let us emphasize that both parametrizations contain the same information about the asymptotic structure of the two-particle vertices, as the \(K_n\) and \(Q_n\) parametrizations can be exactly transformed into each other. For an appropriate choice of numerical frequency grids, both parametrizations are therefore equally valid and differ only in numerical performance. The former approach allows for a more fine-grained adjustment of discrete frequencies to the asymptotic decay of individual classes, while the latter reduces the cost of evoking a two-particle vertex from a summation of up to four classes $K_n$ to loading just a single $Q_n$.

The central observable computed from the pffRG equations is the flowing spin-spin correlation function,
\begin{align}
	\chi^{\mu \nu}_{ij}(i \omega = 0) = \int_{0}^{\infty} d\tau \langle T_{\tau} S^{\mu}_{i}(\tau) S^{\nu}_{j}(0) \rangle \,,
	\label{eq:susceptibility}
\end{align}
where we omit indication of the \(\Lambda\)-dependence for brevity. In all models considered here, the interactions in the Hamiltonian are diagonal and \(\mathrm{SU}(2)\)-symmetric. This leads to spin-spin correlations that are symmetric as well, and we thus define \(\chi_{ij} \equiv \chi^{xx}_{ij} = \chi^{yy}_{ij} = \chi^{zz}_{ij}\).

\begin{figure}[t]
    \centering
    \includegraphics[width=\columnwidth]{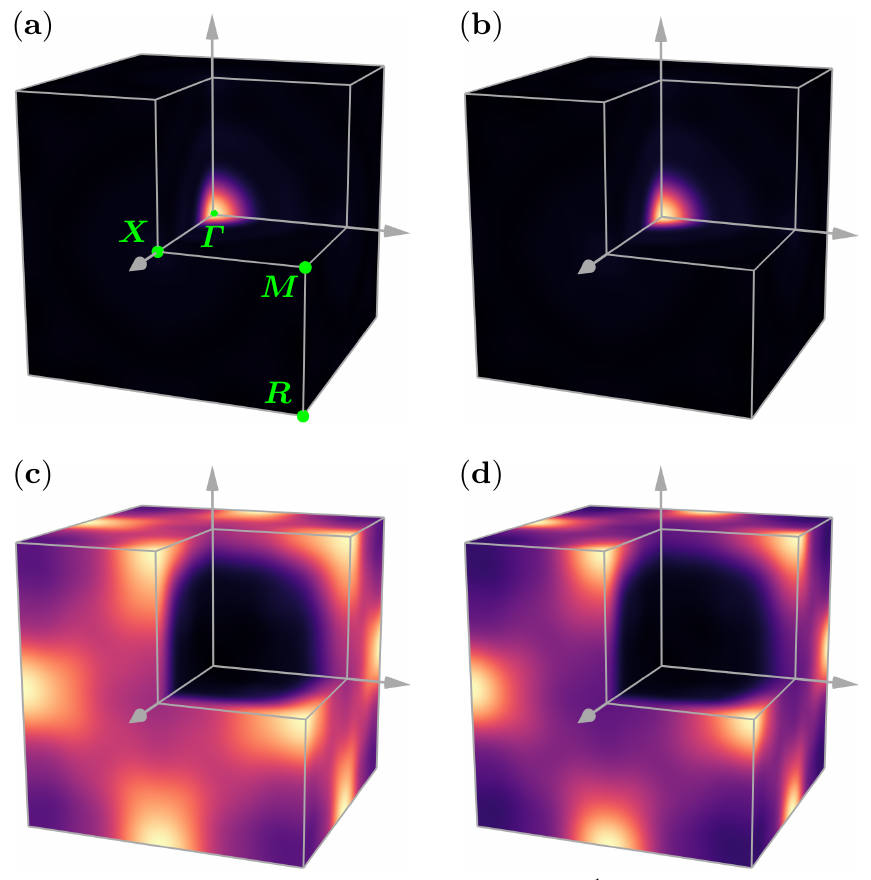}
    \caption{
    {\bf Momentum-resolved structure factors} within the first Brillouin zone of the cubic lattice for (a, b) the ferromagnetic case at \(\Lambda/J = 0.8\) and (c, d) the paramagnetic case at \(\Lambda/J = 0.3\), computed for (a, c) \(\ell = 1\) and (b, d) \(\ell = 3\) using code \#2.
    The ferromagnet shows a sharp peak at the \(\vec{\Gamma}\) point, without visible difference between the two loop orders.
    The putative paramagnet shows a broadened distribution of spectral weight centered around soft maxima at the $\vec{M}$ points in $1\ell$ calculations, while the structure factor peaks more distinctively for $\ell=3$, signalling the onset of magnetic order instead.
    }
    \label{fig:PM_spinstructure_factor_BZ}
\end{figure}

The spin-spin correlations can be used to identify transitions into phases with broken symmetries; there, the flow becomes unstable at some $\Lambdatransition$ and must be stopped. For long-range ordered states, the momentum $\vec{k}$ for which the structure factor
\begin{equation}
    \chi(\vec{k}, i\omega)
    =
    \frac{1}{N_{\text{sites}}}
    \sum_{ij} e^{i \vec{k} \cdot (\vec{R}_i - \vec{R}_j)} \chi_{ij} (i\omega)
\end{equation}
(i.e.\ the Fourier transform of $\chi_{ij}$) is most dominant gives an indication of the emergent magnetic order, as exemplified in Fig.~\ref{fig:PM_spinstructure_factor_BZ}. A smooth flow down to the infrared $\Lambda \to 0$ is, on the other hand, associated with non-magnetic phases, such as spin liquids, dimerized, or plaquette-ordered states.


\section{Results}
\label{sec:results}

\begin{figure}[b]
    \centering
    \includegraphics{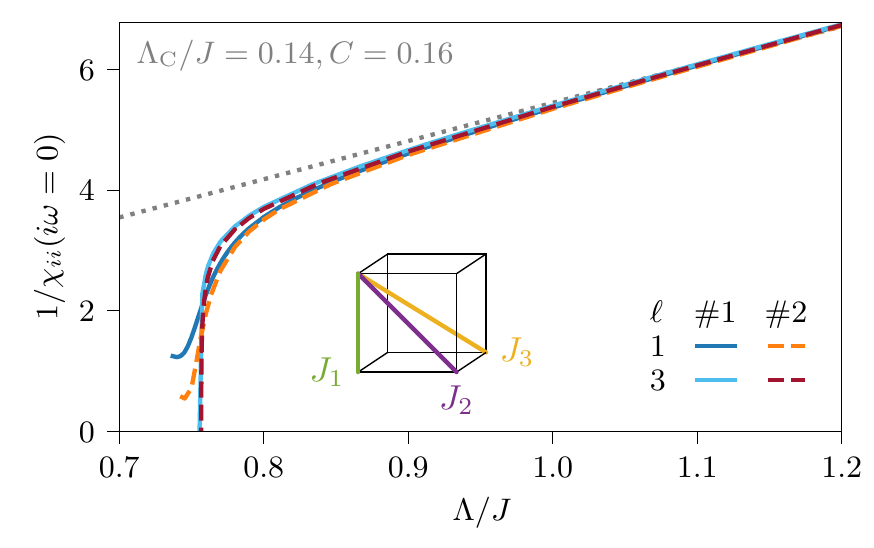}
    \caption{
        {\bf Inverse spin-spin correlation function for the ferromagnet} as a function of \(\Lambda\). Shown here is a comparison of the \(\ell=1\) and \(\ell=3\) flows obtained from both codes. The dotted line is a  
        \(\Lambda^{-1}\) fit [\(\chicw = CJ/(\Lambda - \Lambdacw)\)] to the data at \(\Lambda/J\in\left[1.0, 4.0\right]\). The transition to a ferromagnetically ordered phase is visible as a sharp downturn away from Curie--Weiss behavior.
        {\bf Inset:}
        Definition of the first, second, and third nearest-neighbor interaction, $J_1$ (green), $J_2$ (purple), and $J_3$ (yellow).
        }
    \label{fig:cube_unit_cell}
    \label{fig:ferro_susceptibility_flow}
\end{figure}

In order to benchmark the two codes, we calculate the spin-spin correlations and pseudofermion vertices of an extended Heisenberg model on the cubic lattice with a maximum correlation length $\xi = 5$ in units of the lattice spacing \cite{thoenniss2020}. The corresponding three-dimensional cluster 
contains $N = 515$ sites, small enough to efficiently compare the two codes but large enough to produce the (qualitatively) correct physics. The corresponding Hamiltonian with up to third-neighbor interactions (see inset in Fig.~\ref{fig:cube_unit_cell}) reads
\begin{equation}
	\mathcal{H} =
	J_1 \sum_{\langle ij\rangle} S^{\mu}_i S^{\mu}_j +
	J_2 \sum_{\langle\langle ij\rangle\rangle} S^{\mu}_i S^{\mu}_j +
	J_3 \sum_{\langle\langle\langle ij\rangle\rangle\rangle} S^{\mu}_i S^{\mu}_j
	\,,
	\label{ham_SU2}
\end{equation}
where we fix \(J \equiv \sqrt{J_1^2 + J_2^2 + J_3^2}\) as the unit of energy. We focus on two choices of these interaction parameters to highlight differences between fRG flows in different phases:
\begin{alignat}{3}
J_1 & < 0, & \qquad J_2 & = 0, & \qquad J_3 & = 0,
\label{ferro}
\\
J_1 & > 0, \qquad & J_2/J_1 & = 0.6, \qquad & J_3/J_1 & = 0.25,
\label{para}
\end{alignat}
where Eq.~\eqref{ferro} yields a nearest-neighbor ferromagnet
and the setup of Eq.~\eqref{para} was previously reported to result in a paramagnetic ground state \cite{Iqbal3D}.

Rewriting each spin operator \(S^\mu\) in the Hamiltonian in terms of pseudofermions leads to an expression proportional to
\(
    f^\dagger_{\alpha'} f^{\vphantom{\dagger}}_{\alpha} f^\dagger_{\beta'} f^{\vphantom{\dagger}}_{\beta}
\),
with interactions proportional to
\(
    \sum_\mu \sigma^{\mu}_{\alpha'\alpha} \sigma^{\mu}_{\beta'\beta}
\).
Exploiting this \(\mathrm{SU}(2)\) symmetry (the interactions are diagonal and of equal magnitude in every spin direction), the flowing pseudofermion vertex \(\Gamma\)
(and each of its two-particle reducible parts $\gamma_c$) can be decomposed into a spin component \(\Gamma^s\), proportional to the latter combination of Pauli matrices, and a density component \(\Gamma^d\) proportional to \(\delta_{\alpha'\alpha}\delta_{\beta'\beta}\) \cite{ReutherOrig,BuessenOffDiag}. 
Note that the density component, although initially vanishing for any typical spin model, becomes finite away from the UV limit and is essential for tracking the evolution of {\sl all} symmetry-allowed couplings under the RG flow.

\begin{figure}[t]
    \centering
    \includegraphics{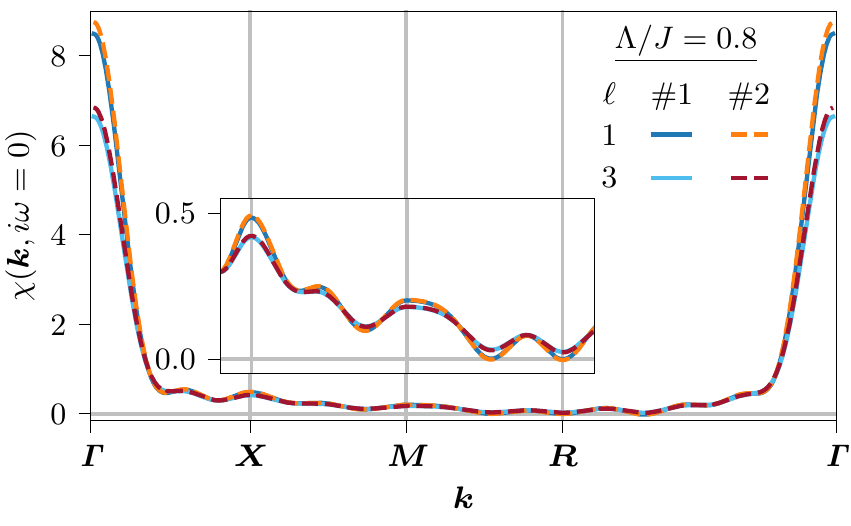}
    \caption{{\bf Structure factor for the ferromagnet} along a high-symmetry path of the cubic lattice Brillouin zone. The results are in excellent agreement between both codes, both for $\ell = 1$ and $\ell = 3$, showing dominant ferromagnetic correlations indicated by a sharp peak around the $\vec{\Gamma}$ point. {\bf Inset:} Zoom into 
    the path segment connecting the $\vec{X}, \vec{M}$, and $\vec{R}$ point.} 
        \label{fig:ferro_path}
\end{figure}

\begin{figure*}
    \centering
    \includegraphics{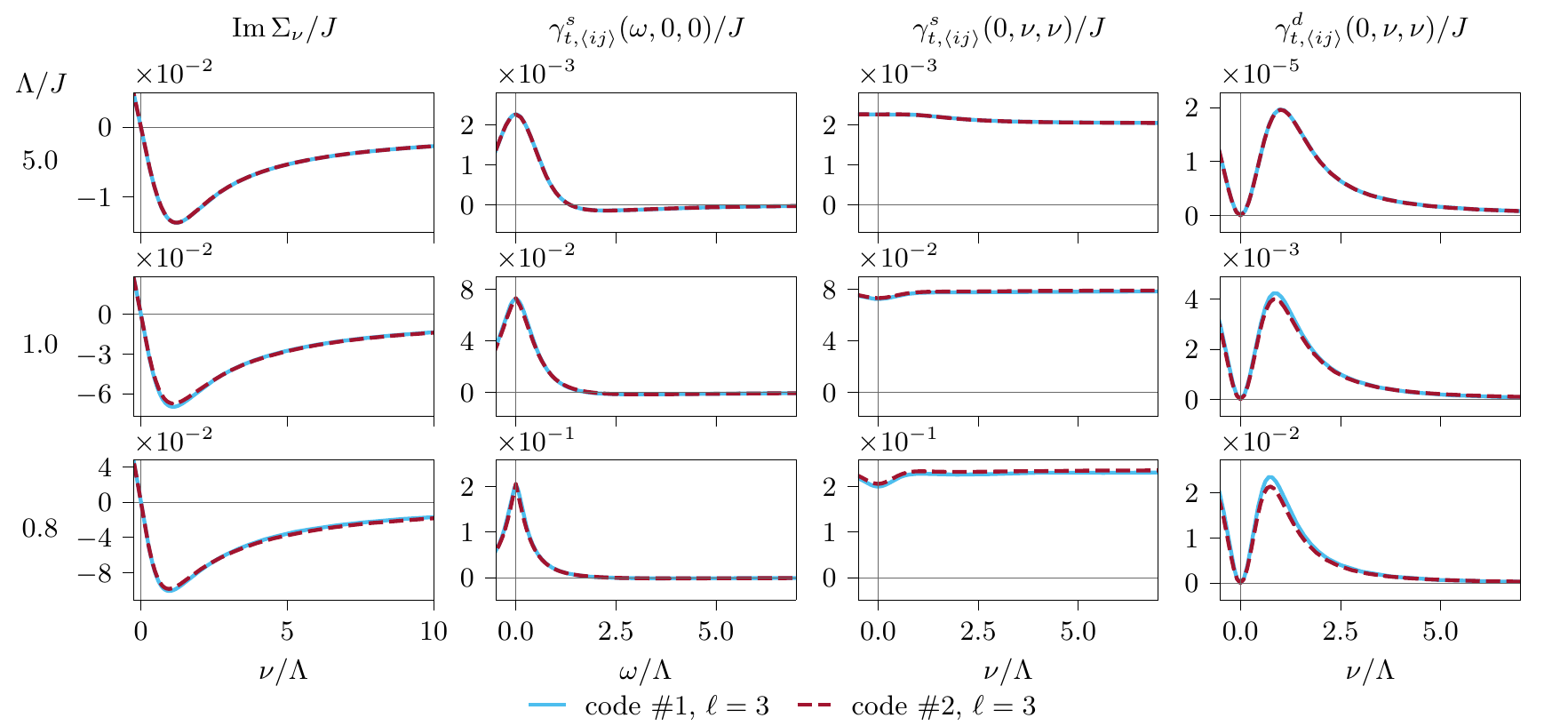}
    \caption{
        {\bf Frequency structure of self-energy and \(t\)-reducible vertex for the ferromagnet} at different values of \(\Lambda / J\) for \(\ell = 3\) flows. The self-energy is purely imaginary and antisymmetric in frequency space, while all vertex components are real and symmetric along the directions plotted here. 
        We show two cuts through the three-dimensional structure of \(\smash{\gamma_{t, \langle ij\rangle}^{\Lambda, \mu} (\omega, \nu, \nu')}\): A cut along the bosonic frequency axis \(\omega\), with both fermionic frequencies set to \(\nu = \nu' = 0\), and a cut with equal fermionic frequencies \(\nu = \nu'\), where the bosonic frequency was set to \(\omega = 0\). The first cut is not shown for \(\gamma_t^d\) as \(\smash{\gamma_{t, \langle ij\rangle}^{d}(\omega, 0, 0) = 0}\) due to symmetry \cite{thoenniss2020,kiese2021}.
        The most prominent structure in the \(t\)-reducible vertex is a peak around zero bosonic frequency \(\omega = 0\) that grows in magnitude and becomes sharper as \(\Lambda\) is decreased. This indicates ferromagnetic correlations that grow stronger as the ordering phase transition is approached. In all components, there is quantitative agreement between the two codes. 
    }
    \label{fig:FM_vertex_flow}
\end{figure*}

\subsection{Ferromagnetic phase}

With pure nearest-neighbor ferromagnetic interactions, the zero-temperature ground state is intuitively expected to be a ferromagnet. Therefore, in the context of pseudofermion fRG, there should be a transition at some finite \(\Lambdatransition > 0\) from a paramagnetic regime at large \(\Lambda > \Lambdatransition\) to the ferromagnetic phase at \(\Lambda < \Lambdatransition\).
Approaching the transition, the spin-spin correlator \(\chi_{ij}\) is expected to diverge, similar to a finite-temperature phase transition.
In this case, a peak will form at the \(\vec{\Gamma}\) point in reciprocal space, as is visible Fig.~\ref{fig:PM_spinstructure_factor_BZ}, since the correlations are uniform and positive in a ferromagnet.
 
Close to the transition, the flow is supposed to visibly deviate from its paramagnetic Curie--Weiss behavior $\chi_{ii} \approx {CJ}/{(\Lambda - \Lambdacw)}$ at large $\Lambda \gg \Lambdatransition$. For this reason, it is convenient to plot the inverse correlator \(1/\chi_{ii}\) as a function of \(\Lambda\) to locate the transition, as shown in Fig.~\ref{fig:ferro_susceptibility_flow}. Here, the $1 / \Lambda$ behavior appears as a straight line with slope \(1/C\) displaced horizontally by \(\Lambdacw/J\) and the transition to the ferromagnetic phase is visible as a sharp turn down to a smaller inverse correlation function at \(\Lambda/J \approx 0.76\). The structure factor at \(\Lambda\) close to \(\Lambdatransition\), shown in Figs.~\ref{fig:PM_spinstructure_factor_BZ} and \ref{fig:ferro_path}, has a single peak at the \(\vec{\Gamma}\) point, signifying an instability towards ferromagnetic order. This, as well as the Curie--Weiss fit parameters, are consistent across both considered loop orders \(\ell = 1, 3\) and both codes, while \(\Lambdatransition\) differs slightly.

Since both implementations obtain the spin-spin correlations by post-processing the vertices, any discrepancy therein originates from differences in the vertices. Therefore, a more detailed examination of the \(1 / \chi_{ii}\)-deviations between the codes for \(\ell = 1\) will follow once the flow of the vertex components has been discussed. Moreover, even if the flows for the $\chi_{ij}$ agree perfectly (as, e.g., in the regime $\Lambda > \Lambdatransition$), discrepancies in the vertices cannot be fully excluded, as post-processing spin-spin correlations from pseudofermion vertex data amounts to integrating a combination of several propagators and the vertex over two frequencies \cite{thoenniss2020}. Hence, this additional step might hide potential differences in the vertex data.

\begin{figure}[t]
    \centering
    \includegraphics{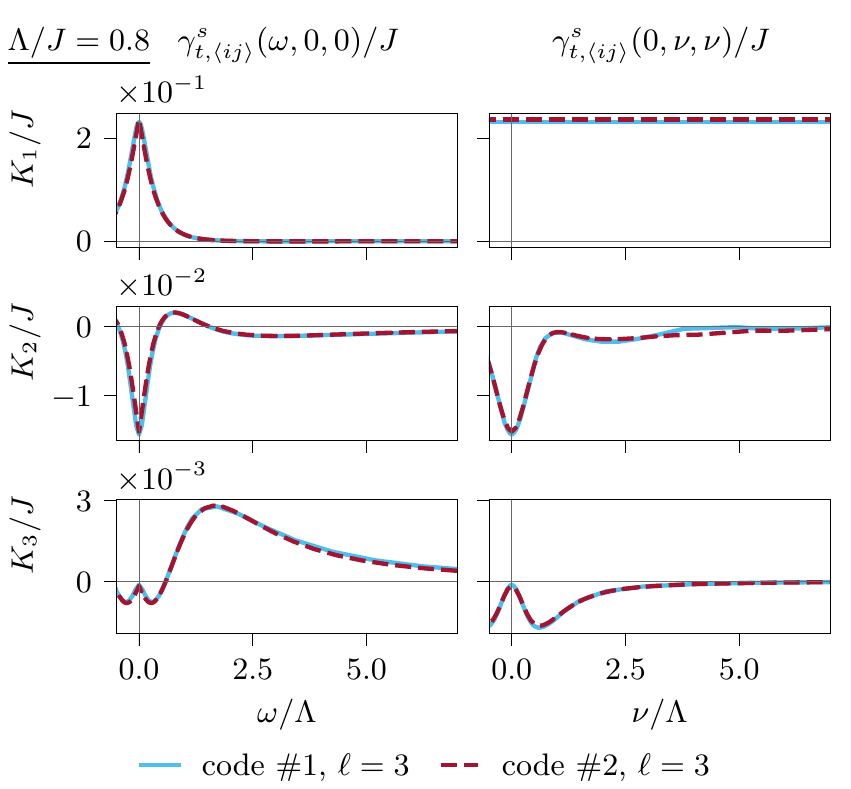}
    \caption{
        {\bf Decomposition of the \(\smash{\gamma_{t, \langle ij\rangle}^{s}(\omega, \nu, \nu')}\) vertex for the ferromagnet} into asymptotic classes \(K_{1,t}, K_{2,t}, K_{3,t}\) (first, second, third row) for the  \(\ell \!=\! 3\) flows
        at \(\Lambda / J \!=\! 0.8\).
        Frequency axes shown here are the same as in Fig.~\ref{fig:FM_vertex_flow}.
        As the flow is close to the ordering phase transition at this value of \(\Lambda\), strong ferromagnetic correlations are present as a peak around \(\omega = 0\) in \(K_{1,t}\). The other classes are at least one order of magnitude smaller.
        In all classes, both codes show quantitative agreement.
    }
    \label{fig:FM_vertex_decomposition}
\end{figure}

To investigate this further, we focus on the \(t\)-reducible vertex \(\gamma_t\) plotted in Fig.~\ref{fig:FM_vertex_flow} at various values of \(\Lambda\): Its spin component \(\gamma_t^{s}\) (second and third column) is responsible for the transition and becomes sharply peaked at small bosonic frequencies \(\omega \approx 0\). Its density component \(\gamma_t^{d}\) (last column) with its extended structures and peaks at non-zero fermionic frequencies \(\nu\) is particularly difficult to resolve and thus most likely to contain numerical artifacts. Comparing \(\gamma_t\), as well as the the self-energy \(\Sigma\) between the codes, we find quantitative agreement also on this very detailed level of inspection.

As outlined in Sec.~\ref{sec:mfRG}, both codes use a decomposition of the reducible vertices \(\gamma_s, \gamma_t, \gamma_u\) into four asymptotic classes each. The decomposition into asymptotic classes $K_n$ is shown for \(\gamma_t^{s}\) at \(\Lambda / J =0.8\) in Fig.~\ref{fig:FM_vertex_decomposition}, where we omit \(K^s_{2',t}\), as it is equal to \(K^s_{2,t}\) by crossing symmetry \cite{thoenniss2020,kiese2021}. Note that, while these vertices can directly be extracted from code \#1, an additional transformation is applied to the $Q_n$ decomposition of code \#2 [see Eq.~\eqref{Qdef}]. The peak in $\gamma^{s}_{t}$ at small bosonic frequencies in Fig.~\ref{fig:FM_vertex_flow} is found to stem from the \(K_1\) contribution, which is an order of magnitude larger than the other classes. In \(K_2\) and \(K_3\), extended structures with multiple maxima and minima exist. It is thus crucial to use a frequency mesh with enough mesh points in an extended region around the origin to control numerical interpolation errors (see Sec.~\ref{sec:technical_aspects}).

Though the codes implement the vertex decomposition differently (see Sec.~\ref{sec:mfRG}) and use different approaches to build appropriate frequency meshes (see \cite{thoenniss2020,kiese2021} for a detailed description), all components of the vertex are consistent with each other. This demonstrates that it is possible to gain control over said interpolation errors by a careful adaptive implementation that places enough mesh points where they are needed.

Since the numerical error incurred by interpolation of the continuous frequency structure from a discrete mesh is particularly relevant whenever sharp structures are present in the vertex, different choices of frequency meshes have strong effects close to phase transitions, where some couplings are expected to diverge. For instance, in the ferromagnetic setup discussed above, the transition was induced by a peak in the spin component of the \(t\)-reducible vertex that grows quickly and starts to diverge, as can be seen in the second column of Fig.~\ref{fig:FM_vertex_flow}. As the transition is approached, this peak progressively becomes sharper and thus more difficult to resolve using discrete meshes. Thus, minor differences in mesh spacing can induce differences in the flow at the transition, though the qualitative, physical results remain unchanged.

\begin{figure}[t]
    \centering
    \includegraphics{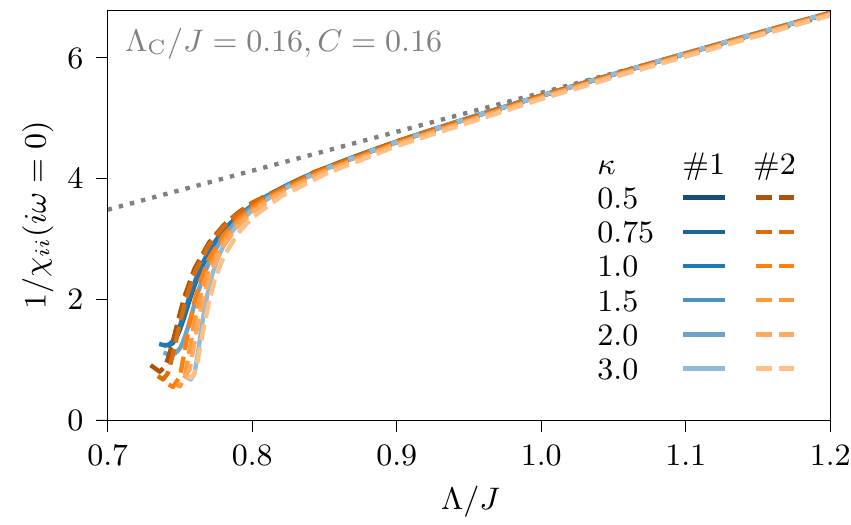}
    \caption{{\bf Flows with rescaled frequency meshes.}
        Comparison of the flow of 
        inverse static on-site spin correlations \(1/\chi_{ii}(i\omega=0)\)
        obtained using frequency meshes with different scaling factors \(\kappa\). The dotted line is a \(\Lambda^{-1}\) fit to the data at \(\Lambda / J \in [1.0, 4.0]\). For all values of \(\kappa\), a transition to a ferromagnet is visible as a sharp turn down. The predicted transition point as well as the slope of \(\chi\) in the region \(\Lambda/J < 0.8\) differs, while the behavior at large \(\Lambda > J\) remains identical.
    }
    \label{fig:ferro_susceptibility_flow_meshspacing}
\end{figure}

To investigate the effects of changes in the mesh spacing explicitly, we compared results obtained from both codes with artificially modified meshes. Both implementations make use of adaptive frequency grids where, during the flow, the mesh spacing is adjusted according to the frequency structure of the vertex. The simplest way to manipulate the meshes is to rescale them by an artificial scaling factor \(\kappa\). In Fig.~\ref{fig:ferro_susceptibility_flow_meshspacing}, we show the effect of such a rescaling on the \(\ell=1\) flow from Fig.~\ref{fig:ferro_susceptibility_flow}. Above \(\Lambda /J \approx 0.8\), all frequency structures in the vertex are fairly broad and easy to resolve. Consequently, rescaling the frequency grid has little effect and values \(\kappa = 0.5 \ldots 3.0\) result in the same flow and also the same Curie--Weiss fit parameters. Below that point, the flows differ more and more as structures become sharper and ultimately predict slightly different transition points \(\Lambdatransition/J\). Nevertheless, all flows predict a transition to the same ferromagnetic phase, which can be identified by a peak in the structure factor at the \(\vec{\Gamma}\) point.

\begin{figure}[b]
    \centering
    \includegraphics{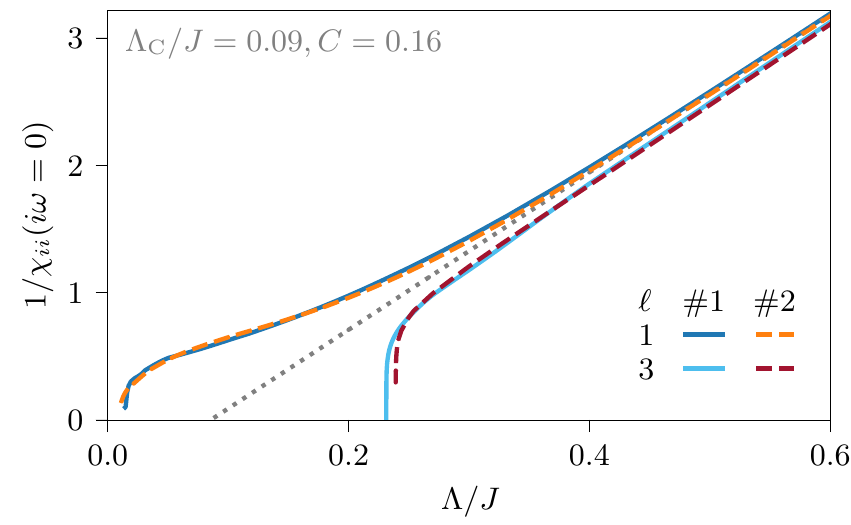}
    \caption{
        {\bf Inverse spin-spin correlation function for the putative paramagnet} as a function of \(\Lambda\). Shown here is a comparison of the \(\ell=1\) and \(\ell=3\) flow obtained from both codes.
        The dotted line is a fit of 
        a \(\Lambda^{-1}\) power law
        to the data at \(\Lambda / J \in [1.0, 4.0]\).
        For \(\Lambda/J \geq 0.5\), the $\Lambda^{-1}$ behavior is followed almost perfectly.
        At smaller \(\Lambda/J\), the \(\ell=1\) and \(\ell=3\) flows disagree: The \(\ell=1\) curve smoothly approaches $\Lambda=0$ (staying above the power law), indicating antiferromagnetic correlations. By contrast, the \(\ell=3\) curve displays a downward cusp, similar to Fig.~\ref{fig:ferro_susceptibility_flow}, and thus predicts an ordered state.
    }
    \label{fig:PM_susceptibility_flow_comparison}
\end{figure}

\begin{figure*}[h!]
    \centering
    \includegraphics{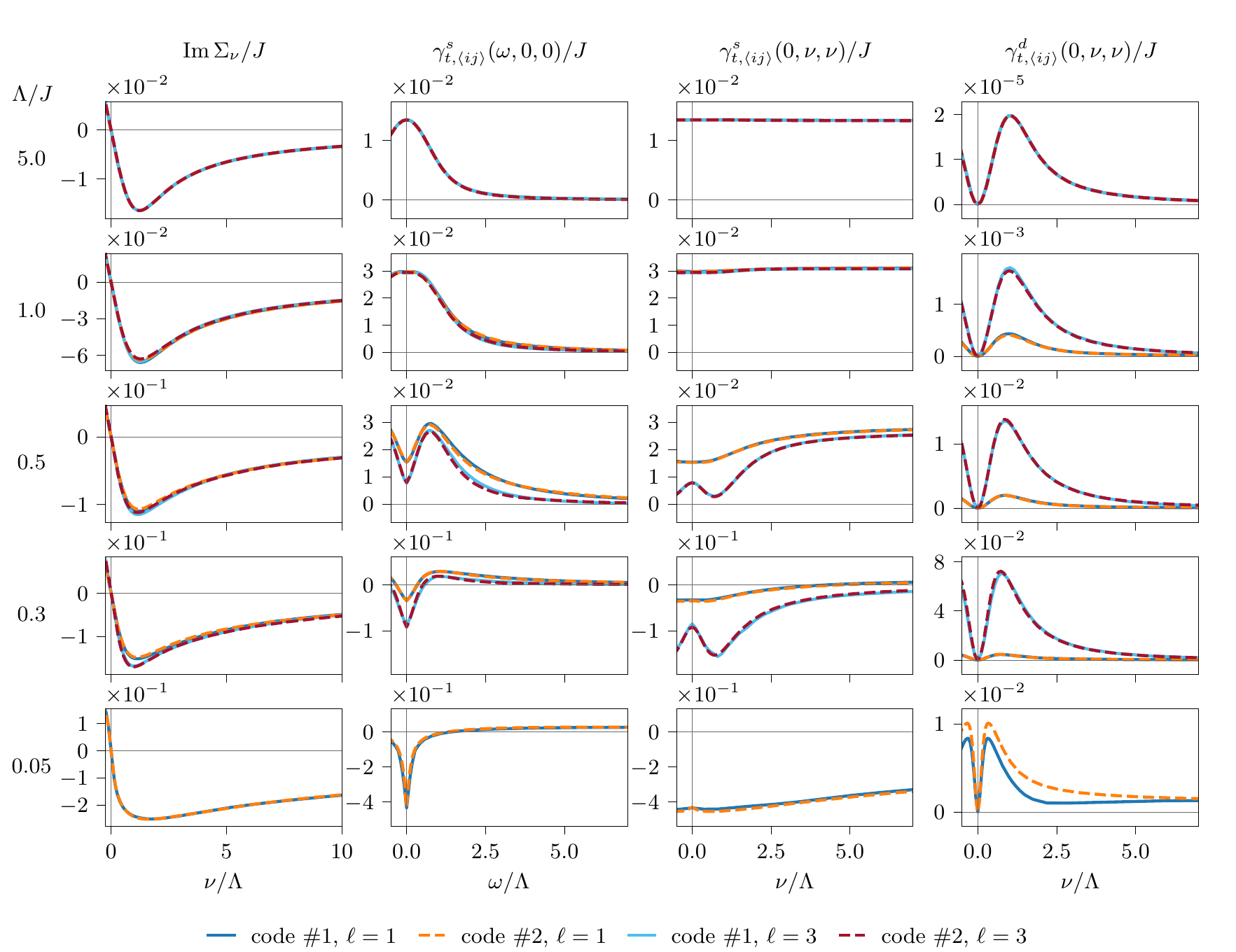}
    \caption{
        {\bf Frequency structure of self-energy and \(t\)-reducible vertex for the putative paramagnet} at different values of \(\Lambda / J\) for \(\ell = 1\) and \(3\) flows.
        As the \(\ell = 3\) flow diverges at \(\Lambda / J \approx 0.24\), only \(\ell = 1\) is shown at \(\Lambda / J = 0.05\).
        The same cuts through the three-dimensional frequency structure of the vertices are shown as in Fig.~\ref{fig:FM_vertex_flow}.
        Again, a peak in the \(\gamma^{s}_{t, \langle ij\rangle}\) component (second column) indicates strong correlations that become stronger as \(\Lambda\) is further decreased. In contrast to the ferromagnetic case, this peak is negative, indicative of antiferromagnetic correlations, and there is a sizeable contribution of $\gamma^s_t$ for nonzero fermionic frequencies \(\nu, \nu'\) (third column), particularly for \(\ell = 3\).
    }
    \label{fig:PM_vertex_flow}
\end{figure*}

\begin{figure}
    \centering
    \includegraphics{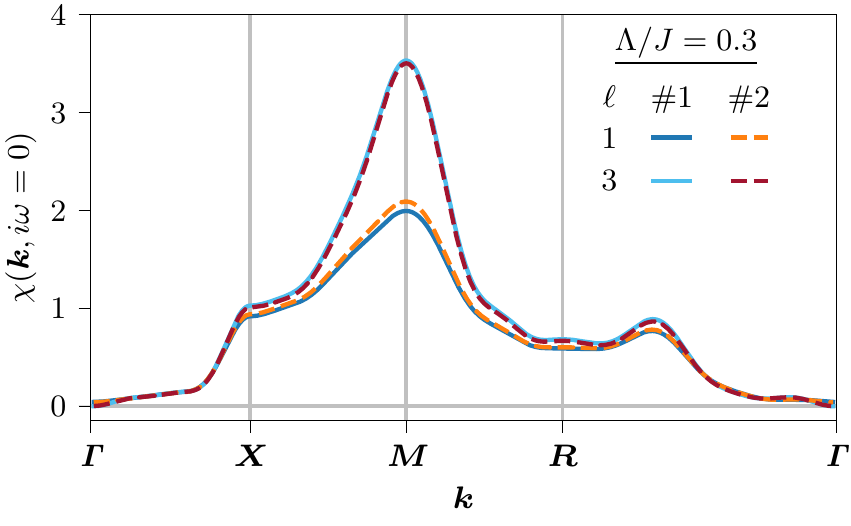}
    \caption{{\bf Structure factor for the paramagnetic setup} along a high-symmetry path of the cubic lattice Brillouin zone. The results are in good agreement between both codes, both for $\ell \!=\! 1$ and $\ell \!=\! 3$, showing that correlations are strongest around the $\vec{M}$ point. Here, the peak sharpens with increasing loop order, and the $3\ell$ flow predicts enhanced long-range correlations.}
    \label{fig:para_path}
\end{figure}

\subsection{Paramagnetic phase}

For the second set of parameters, Eq.~\eqref{para}, all interactions up to the third neighbor are antiferromagnetic. Consistent with prior work using one-loop fRG \cite{Iqbal3D}, both codes find a paramagnetic ground state for \(\ell=1\), indicated by a smooth and regular flow down to $\Lambda=0$ in Fig.~\ref{fig:PM_susceptibility_flow_comparison}.

Remarkably, the \(\ell=3\) data predicts a qualitatively different phase: There is a divergence in the spin correlations at $\Lambdatransition / J \approx 0.24$, indicating an ordering transition at a scale roughly three times lower than for the ferromagnetic ordering instability discussed in the previous section. Such a reduced ordering scale is not unexpected for an exchange-frustrated spin system when compared to an unfrustrated one, but sometimes hard to establish. 

Probing the structure factor in the vicinity of the divergence reveals a strong enhancement of magnetic correlations compared to the $\ell = 1$ flow, as indicated by sharpened Bragg peaks around the $\vec{M} = (0, \pi, \pi)$ points in Figs.~\ref{fig:PM_spinstructure_factor_BZ} and \ref{fig:para_path}. These correspond to antiferromagnetic correlations between planes orthogonal to the vector connecting the second nearest-neighbors along diagonals of the faces in the cubic unit cell (shown in purple in Fig.~\ref{fig:cube_unit_cell}). Our result is consistent with earlier observations of long-range $(0, \pi, \pi)$ order neighboring the paramagnetic phase \cite{Iqbal3D}. Yet, the mfRG flows obtained from both codes suggest a rather strong modification of the respective phase boundaries as the coupling parameters investigated here were previously predicted to be deep in the non-magnetic regime.

\begin{figure}[h!]
    \centering
    \includegraphics{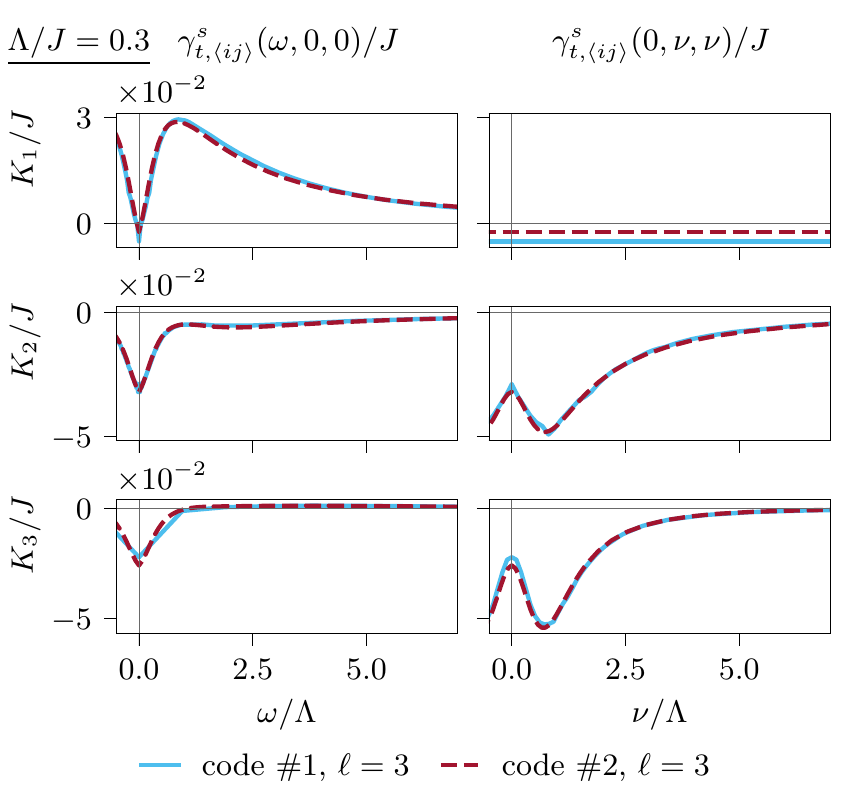}
    \caption{
        {\bf Decomposition of the \(\smash{\gamma_{t, \langle ij\rangle}^{s}(\omega, \nu, \nu')}\) vertex in the paramagnetic setup}
        as in Fig.~\ref{fig:FM_vertex_decomposition},
        for the  \(\ell = 3\) flows at \(\Lambda / J = 0.3\).
        Here, all asymptotic classes are of the same order of magnitude, and structures with multiple peaks are present along the fermionic frequency cut (second column).
    }
    \label{fig:PM_vertex_decomposition}
\end{figure}

In the vertex (see Fig.~\ref{fig:PM_vertex_flow}) and self-energy, 
there is again very good quantitative agreement between both codes. At \(\Lambda/J = 0.05\), small quantitative differences between code \#1 and \#2 appear in the density component \(\gamma_t^d\) of the \(t\)-reducible vertex, consistent with the earlier remark that it is the most difficult component to resolve well.

The \(\ell = 1\) and \(\ell = 3\) flows are very similar down to \(\Lambda/J \geq 1\). Contributions of \(\ell > 1\) terms become significant at \(\Lambda/J \approx 1\) and eventually lead to 
an ordering instability 
induced by a peak in the \(\smash{\gamma_{t}^{s}}\) component that diverges 
at \(\Lambda / J \approx 0.24\).
In contrast to the ferromagnetic case, this peak is negative, indicating anti-correlation. Along the fermionic \(\nu\) frequency axis, the vertex shows an extended structure with multiple peaks of similar magnitude to the one on the bosonic axis.
Since the \(K_1\) class has no fermionic frequency, this means that, remarkably, other classes reach an order of magnitude comparable to \(K_1\), as shown explicitly in Fig.~\ref{fig:PM_vertex_decomposition}.
Consequently, vertex structures along fermionic frequency axes, in contrast to the ferromagnetic transition, become sizeable.
It is therefore crucial to resolve the full three-dimensional frequency structure in \(K_3\). Though numerically expensive, a large number of mesh points is necessary to ensure sufficient accuracy, as inadequate resolution of features along the fermionic frequency axes can strongly affect the fRG flow. This is even more important for multiloop flows, where interpolation errors might accumulate during the iteration over loop orders.


\section{Technical aspects}
\label{sec:technical_aspects}

To conclude our benchmark calculations, we discuss some of the particularly relevant technical aspects (see Tab.~\ref{tab:algorithms}) which are needed to obtain confidence that we have sufficient degree of control over numerical errors. In doing so, we will also connect to the existing literature and scrutinize some of the algorithmic approaches which are routinely employed in the pffRG community.

\begin{table*}[h!]
    \centering
    \begin{tabular}{r|c|c}
    	\hline\hline	
        \rule{0pt}{12pt} & Code \#1 & Code \#2
        \\[1mm]\hline
        \rule{0pt}{12pt}  Vertex decomposition
            & \(K_1, K_2, K_3\)
            & \(Q_1, Q_2, Q_3\)
        \\[1mm]
        Frequency mesh
            & adaptive linear and algebraic
            & adaptive linear and logarithmic
        \\[1mm]
        Integration rule
            & adaptive 21-point Gauss--Kronrod rule
            & adaptive Simpson rule + Richardson extrapolation
        \\[1mm]
        ODE solver
            & 5th order Cash--Carp
            & 3rd order Bogacki--Shampine
        \\[1mm]\hline\hline
    \end{tabular}
    \caption{
        {\bf Technical summary} of the algorithmic choices in code \#1 and \#2.
    }
    \label{tab:algorithms}
\end{table*}

\subsection{Frequency grids}

Both the self-energy and two-particle vertices are functions of Matsubara frequencies, which are continuous in the zero-temperature limit. A numerical implementation has to sample these functions on a finite grid and interpolate their values inbetween the sampling points. In many previous works (see e.g.\ Refs.~\cite{ReutherOrig,KieseSpinValley,BuessenThesis}), the {\sl same} frequency grid was chosen for the self-energy and all reducible vertices, usually featuring logarithmically increasing distances between adjacent grid points starting from some small but finite frequency. The intention behind such a choice of frequencies was to resolve the structure around zero frequency with high accuracy while coarse-graining high-frequency tails. Moreover, each vertex component was parametrized in terms of the three bosonic transfer frequencies, instead of the channel-specific mixed bosonic-fermionic frequency treatment utilized by codes \#1 and \#2. 

Although most of the structure of the two-particle vertex is indeed centered around zero frequency, its precise extent strongly depends on the cutoff scale $\Lambda$ (see, e.g., Figs.~\ref{fig:FM_vertex_flow} and \ref{fig:PM_vertex_flow}) and a {\sl static} frequency grid will therefore fail to faithfully resolve the evolution of frequency structures under the fRG flow. Furthermore, multipeak structures that are present in several vertex components will in general not be captured by logarithmic sampling.

To address both shortcomings, codes \#1 and \#2 introduce hybrid frequency meshes using linear spacing around zero frequency augmented by an algebraic (code \#1) or logarithmic (code \#2) part to capture the high-frequency behavior in the asymptotic classes $K_n$ or $Q_n$. The parameters of these meshes are then independently rescaled for different vertex components making use of sophisticated scanning routines (see \cite{thoenniss2020,kiese2021} for further details).

\subsection{Evaluation of bubble integrals}

\begin{figure}[b]
    \centering
    \includegraphics[width=\linewidth]{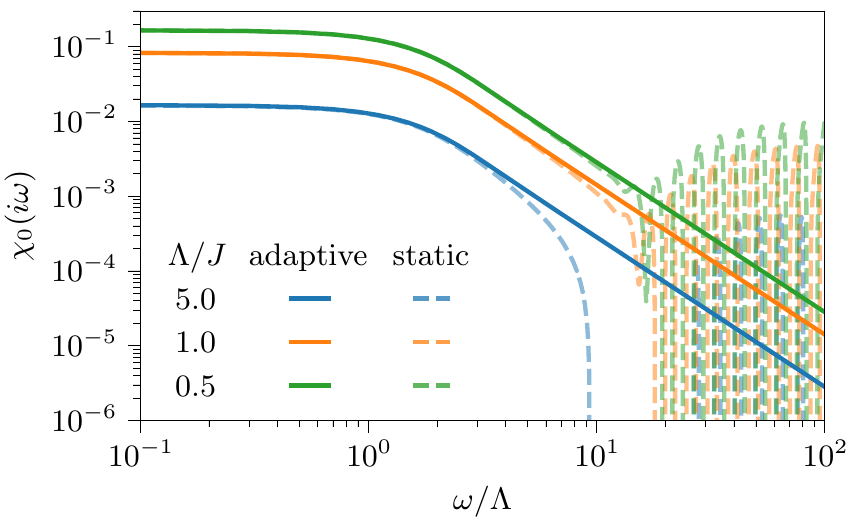}
    \caption{{\bf Evaluation of bubble integrals.} Comparison of the bare susceptibility $\chi^{\Lambda}_{0}(\omega) = \frac{1}{4 \pi} \int d\nu G^{\Lambda}_{0}(\nu + \frac{\omega}{2}) G^{\Lambda}_{0}(\nu - \frac{\omega}{2})$ obtained numerically via adaptive and static quadrature. The adaptive method utilizes the Simpson rule, while the static method applies a trapezoidal rule to a fixed logarithmic frequency discretization (see main text for more details). For frequencies larger than the scale set by the cutoff $\Lambda$, the non-adaptive integration becomes unstable and is plagued by rapid oscillations. By contrast, the adaptive routine yields stable results even beyond the small frequency regime and is therefore crucial to obtain accurate results for the vertex functions and their asymptotic behavior.}
    \label{fig:bubble_integration}
\end{figure}

Having fixed the frequency discretization, the evaluation of frequency integrals in loop and bubble functions necessitates the use of a quadrature rule. In earlier implementations, a trapezoidal quadrature was used, with integration points coinciding with the frequency mesh of the vertex. As discussed above, this procedure yields good resolution around the origin of the integration variable. For $1\ell$ calculations, the bubble function consists of a single-scale and a full propagator, the former being more strongly peaked than the latter. As the integration variable was usually shifted such that the origin coincided with the more important pole of the single-scale propagator, at least the dominant contribution was accounted for in previous implementations. 

In higher loops, however, both propagators enter the bubble on equal footing, necessitating {\sl adaptive} routines to deal with the enriched frequency structure. This is illustrated in Fig.~\ref{fig:bubble_integration}, where we compare the results of integrating the bare susceptibility
\begin{equation*}
    \chi^{\Lambda}_{0}(\omega)
    =
    \frac{1}{4 \pi} \int d\nu \,
    G^{\Lambda}_{0}(\nu + \tfrac{\omega}{2}) \, G^{\Lambda}_{0}(\nu - \tfrac{\omega}{2})
    \,,
\end{equation*}
i.e., the simplest bubble-like integral encountered during the fRG flow. 
Using trapezoidal quadrature over a fixed set of $60$ logarithmically distributed integration points between $\nu_{\textrm{min}} = 10^{-3} J$ and $\nu_{\textrm{max}} = 250 J$, we find strong deviations for frequencies $\omega / \Lambda \gtrsim 1 \sim 10$ compared to the results produced with the adaptive routine of code \#2 (see Ref.~\cite{kiese2021} for further details). Moreover, at small cutoffs $\Lambda / J \lesssim 1$, the non-adaptive result is plagued by rapid oscillations, rendering it numerically unstable and thus inapplicable. Analytically, an asymptotic falloff with a power law \(\omega^{-2}\) is expected, and this is reproduced perfectly by the adaptive integrator.

We emphasize that the test case considered here merely constitutes the simplest version of a bubble-like integral computed within the pffRG flow. In general, the propagators in bubble functions are dressed with self-energy insertions and additionally contracted with two-frequency dependent vertices. One should therefore expect even larger numerical errors for full fRG calculations that utilize non-adaptive quadrature.

\begin{figure*}[h!]
    \centering
    \includegraphics{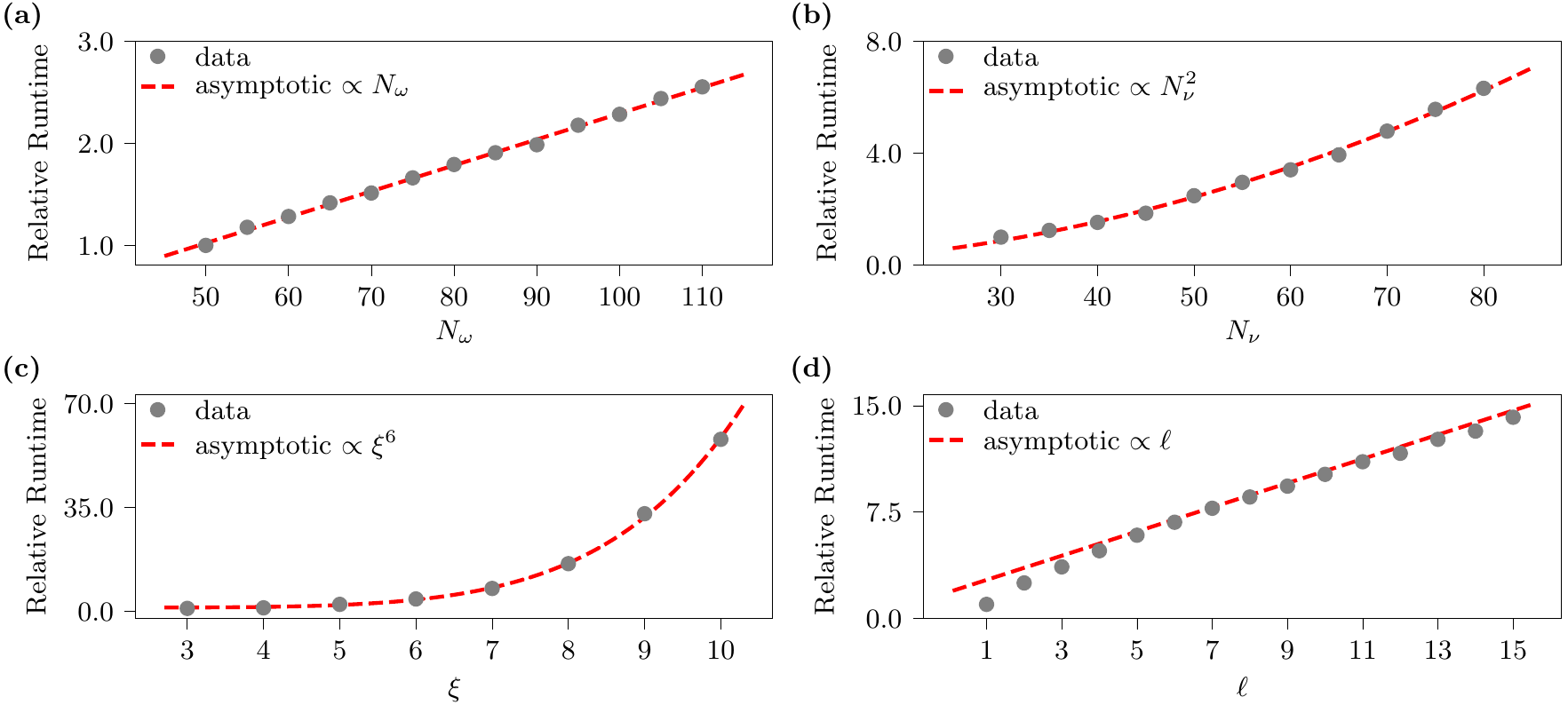}
    \caption{{\bf Scaling of relative runtime with numerical parameters.} Median computational runtime of 60 samples of a single calculation of the right-hand side of the flow equation for $\Lambda / J = 1$ relative to the runtime of the fastest computation in each series. Calculations start from a parquet solution to make the code integrate over non-trivial frequency structures. The numerical parameters for all plots are fixed to $N_\omega = 50$, $N_\nu = 30$, $\xi = 4$ and $\ell = 1$, if not varied. The asymptotic behavior expected analytically is achieved in all cases (dashed red lines).}
    \label{fig:scaling}
\end{figure*}

\subsection{Flow integration}

The integration of the RG flow can, in principle, be performed using any standard solver for ordinary differential equations. While earlier works used an Euler scheme with decreasing step-sizes (see, e.g., Ref.~\cite{BuessenThesis}), we employ higher-order solvers in the Runge--Kutta family with adaptive step-size control to achieve maximum accuracy while being numerically efficient to operate. It is of particular importance to implement an error-controlling method near ordering instabilities such as the ferromagnetic setup in Section \ref{sec:results}, as otherwise numerical errors may become unacceptably large even at scales \(\Lambda \approx J\).

\subsection{Initial condition}

The final ingredient to set up the pffRG flow is an appropriate initial condition. In the UV limit $\Lambda \to \infty$, the pseudofermion vertex is given by the bare spin coupling, which, in numerical calculations, is naturally implemented by using $J$ as the initial condition at a large but finite value of $\Lambda$.
The mfRG flow will, by construction, reproduce a solution to the parquet equations \cite{Kugler_1,Kugler_2,Kugler_3}, given an initial condition consistent with them. We therefore solve the regularized parquet equations iteratively for an initial scale \(\Lambda / J = 5\) and use the resulting self-energy and reducible vertices as a dynamic, i.e., frequency-dependent starting point for the fRG flow \cite{thoenniss2020}.

\subsection{Scaling analysis}

Most of the runtime needed to evaluate the right-hand side of the flow equations is spent calculating the derivative of the high-dimensional two-particle vertex as given in Eq.~\eqref{eq:bubble_function}. In comparison, the computation time spent for the self-energy derivative of Eq.~\eqref{self} is negligible. Consequently, the (asymptotic) computational complexity is given by
\[
	\mathcal{O}\left(N_\xi^2 \times N_{\omega}^{\vphantom{2}} N_\nu^2 \times \ell \right) \,,
\]
where $N_\xi$ is the number of (symmetry reduced \cite{thoenniss2020,kiese2021}) lattice sites, $N_\omega$ ($N_\nu$) the number of bosonic (fermionic) frequencies, and $\ell$ denotes the number of loops. The total number of sites, in turn, is expected to follow a $\mathcal{O}(\xi^d)$ dependence, where $\xi$ is the maximal correlation length considered and $d$ is the spatial dimensionality of the underlying lattice, with $d=3$ for the simple cubic lattice at hand.

\begin{table}[t]
    \centering
    \begin{tabular}{c|c}
    	\hline\hline	
        \rule{0pt}{12pt} max. correlation length $\xi$ & no. flow equations
        \\[1mm]\hline
        \rule{-3.0pt}{12pt} 3
            & \phantom{00}9 183 600
        \\[1mm]
        5
            & \phantom{0}24 795 720
        \\[1mm]
        7
            & \phantom{0}53 264 880
        \\[1mm]
        9
            & 101 019 600
        \\[1mm]
        11
            & 167 141 520
        \\[1mm]
        13
            & 258 059 160
        \\[1mm]\hline\hline
    \end{tabular}
    \caption{
        {\bf Number of (symmetry reduced) vertex flow equations} for Heisenberg models on the cubic lattice as a function of the maximum correlation length $\xi$. The number of positive frequencies is fixed to $60$ ($50$) for the bosonic (fermionic) Matsubara axis.
    }
    \label{tab:no_equations}
\end{table}

To demonstrate that we indeed reach this asymptotic algorithmic scaling also in numerical implementations we show, in Fig.~\ref{fig:scaling}, the median runtime data for $60$ evaluations of the right-hand side of the fRG equations obtained using code \#2. For the number of bosonic and fermionic frequencies, the expected linear and quadratic behavior, respectively, is achieved over the whole parameter range. Note that, due to the adaptive integration and parallelization used, slight deviations from the theoretical scaling are to be expected.
Similarly, the scaling in the maximal correlation length $\xi$ is achieved for the whole parameter range.
In the number of loops, the linear scaling sets in at $\ell=5$, while for smaller $\ell$ a steeper slope is found. We attribute this behavior to the contributions of higher loops becoming successively smaller, leading to faster converging adaptive loop integrals for given absolute and relative tolerances. That way, the initial overhead of computing two (three) loop corrections, which require twice (thrice) the number of integrals to be evaluated compared to $\ell = 1$, diminishes with increasing loop number and the analytically expected scaling, linear in $\ell$, is recovered.

As a final remark,
we mention that the number of vertex flow equations, another measure of algorithmic complexity, grows rapidly 
as one increases the maximal correlation length considered for a given lattice model. This is summarized in Tab.~\ref{tab:no_equations}.


\section{Conclusions}

We benchmarked two state-of-the-art codes for solving pseudofermion functional renormalization group equations. Our analysis considered both physical observables, i.e.\ spin-spin correlation functions and structure factors, as well as fermionic vertex functions (self-energy and two-particle vertex) for ferro- and antiferromagnetic models on the simple cubic lattice.

For the nearest-neighbor ferromagnet, both codes were in quantitative agreement at least until $\Lambda / J \gtrsim 0.76$, where they consistently predicted a breakdown of the RG flow, indicated by a sharp peak (for $\ell = 1$) or a divergence (for $\ell = 3$) in the spin-spin correlations. The energy scale $\Lambdatransition$ associated with this numerical instability slightly differed, which necessitated an in-depth comparison of the influence of the numerical frequency grid on the obtained results. We found that both fRG solvers, due to the emergence of a singular peak in the $t$ reducible vertex functions, become sensitive to the precise mesh spacing and thus predict marginally different critical scales, although the physical conclusion drawn from the RG flow, i.e.\ the onset of long-range ferromagnetic order, remains the same.

For the antiferromagnetic setup, the $\ell = 1$ results obtained by both codes were in agreement with one another and previous studies \cite{Iqbal3D}, predicting a paramagnetic state, signified by a regular RG flow down to the infrared. For $\ell = 3$, similar numerical agreement between the two codes was found. However, the physical results changed qualitatively: the flow of the spin-spin correlator diverged around $\Lambda / J \approx 0.24$, accompanied by sharp Bragg peaks at the $\vec{M}$ points indicating the formation of antiferromagnetic order at low temperatures. This reinstantiates the importance of including higher loop corrections in pffRG in order to avoid overestimating the extent of paramagnetic phases and to obtain more accurate predictions of ground states in frustrated quantum magnets.

We also elaborated on the importance of employing adaptive numerical algorithms to obtain robust results at all stages of the flow. More explicitly, there are extended structures with multiple peaks in the three-dimensional frequency dependence of several vertex components. As these structures are sizable, it is crucial to resolve them in an accurate manner. We found fixed logarithmic frequencies to be insufficient for structures not centered at zero frequency, and rely instead on adaptive frequency meshes that have been specifically optimized for pffRG vertices. Furthermore, we demonstrated
that the commonly employed quadrature of a trapezoidal rule over a static, logarithmic mesh fails to produce the analytically expected behavior of bare bubble integrations at large frequencies. It is thus unsuitable for providing the essential Matsubara integrals for error-controlled fRG flows.
By contrast, the implementations presented and benchmarked here solve these problems by using highly-accurate, yet efficient adaptive routines (see Tab.~\ref{tab:algorithms}).
We thus believe that, moving forward, they will be widely used for unbiased calculations of (multiloop) ground-state phase diagrams of frustrated magnets from pffRG.


\begin{acknowledgement}
\textit{Acknowledgements}---%
We thank L.~Gresista, Y.~Iqbal, M.~Punk, and J.~Reuther for useful and stimulating discussions and J.~Thoenniss for his pioneering contribution to setting up the Munich multiloop pffRG code.
The Cologne group gratefully acknowledges partial support from the Deutsche Forschungsgemeinschaft (DFG) -- Projektnummer 277146847 -- SFB 1238 (project C02),
the Munich group from DFG under Germany's Excellence Strategy EXC-2111 (Project No. 390814868), the W\"urzburg group from DFG through Project-ID 258499086-SFB 1170 and the W\"urzburg-Dresden Cluster of Excellence on Complexity and Topology in Quantum Matter -- ct.qmat Project-ID 390858490-EXC 2147, and F.B.K.\ from the Alexander von Humboldt Foundation through a Feodor Lynen Fellowship.
The numerical simulations were performed on the JURECA Booster and JUWELS cluster at the Forschungszentrum Juelich, the SuperMUC cluster and Linux clusters at the Leibniz Supercomputing Centre, as well as the CHEOPS cluster at RRZK Cologne.
This research is also part of the Munich Quantum Valley, which is supported by the Bavarian state government with funds from the Hightech Agenda Bayern Plus.
\end{acknowledgement}
\bibliography{mfRG}
\clearpage


\end{document}